%% file: gor_main.tex
\documentclass[10pt]{svproc}
\usepackage{amsmath,amsfonts}
\usepackage{graphicx}
\usepackage{multicol}
\usepackage[bottom]{footmisc}
\usepackage{hyperref}
\usepackage{booktabs}
\usepackage{multirow}
\usepackage{tabularx}
\usepackage{color}

\begin{document}

\title{Clustering scientific publications: lessons learned through experiments with a real citation network}
\titlerunning{Clustering scientific publications}
\author{Vu Thi Huong\inst{1,2} \and Thorsten Koch\inst{1,3}}
\institute{Digital Data and Information for Society, Science, and Culture,
		Zuse Institute Berlin, 14195 Berlin, Germany, \email{huong.vu@zib.de}
        \and
        Institute of Mathematics, Vietnam Academy of Science and Technology, 10072 Hanoi, Vietnam, \email{vthuong@math.ac.vn}
        \and
        Software and Algorithms for Discrete Optimization, Technische Universität Berlin, 10623 Berlin, Germany, \email{koch@zib.de}
        }
\maketitle
\begin{quote}
\noindent {\bf Abstract.} 
Clustering scientific publications can reveal underlying research structures within bibliographic databases. Graph-based clustering methods, such as spectral, Louvain, and Leiden algorithms, are frequently utilized due to their capacity to effectively model citation networks. However, their performance may degrade when applied to real-world data. This study evaluates the performance of these clustering algorithms on a citation graph comprising approx. 700,000 papers and 4.6 million citations extracted from Web of Science. 
The results show that while scalable methods like Louvain and Leiden perform efficiently, their default settings often yield poor partitioning. Meaningful outcomes require careful parameter tuning, especially for large networks with uneven structures, including a dense core and loosely connected papers. These findings highlight practical lessons about the challenges of large-scale data, method selection and tuning based on specific structures of bibliometric clustering tasks.


\end{quote}
\section{Introduction}

The exponential growth of scientific literature across all disciplines has made it increasingly difficult for researchers and institutions to keep track of relevant publications. To address this challenge, automated methods for analyzing and organizing scientific articles have become essential. One important approach is clustering: grouping publications based on similarity in content, citation behavior, or other structural features. These clusters can reveal topical communities or emerging fields.

Clustering methods used for scientific publications typically fall into one of three categories: graph-based methods that exploit citation or co-authorship structures; vector-space approaches based on textual features; and neural embedding methods that learn latent representations. Among these, graph-based approaches are particularly appealing when citation data is available, as they are sparse and directly model the scholarly ecosystem as a network of documents with inter-dependencies.

In citation networks, nodes represent articles and directed edges represent citation links. This natural graph representation opens the door to using network community detection algorithms that aim to identify densely connected subgraphs, interpreted as research communities or topic clusters. Two prominent examples are the \textit{Louvain} (\cite{Blondel_2008}) and \textit{Leiden} (\cite{Traag_2019}) algorithms, which optimize a modularity objective to partition the graph into separated clusters. Another major class is \textit{spectral clustering} (\cite{Ng_2002}), which leverages linear algebraic properties of the graph Laplacian to embed nodes into a low-dimensional space and apply standard clustering there.

Despite their popularity, these methods were often evaluated in synthetic or small real-world networks with idealized properties. Applying them to massive, noisy, and sparsely connected citation graphs from real scientific databases remains a challenge. These networks tend to have skewed degree distributions, disconnected components, citation biases, or field-dependent behaviors that can violate key algorithmic assumptions or degrade performance.

In this work, we evaluate the performance of the spectral, Louvain and Leiden clustering methods on a real citation graph extracted from the Web of Science~\cite{webofscience} (WoS). The dataset includes 698,135 articles and 4,590,190 citations, focusing on two subject areas: Mathematics and Operations Research \& Management Science. Our goal is to understand how these methods behave in practice: when they fail or succeed, what types of community they detect, and how well they align with WoS labels. Although the findings are domain-specific, the insights can be applied more broadly to bibliometric research and scholarly data mining.

\section{A citation network from the Web of Science}
\label{sec:data}
The network used in this study was extracted from the Web of Science database with two subject areas: {Mathematics} (M) and {Operations Research} \& {Management Science} (OR\&MS); see~\cite{Huong_Ida_Koch}. The network was represented as a graph with nodes representing publications and directed edges denoting citations. 

  For the convenience of the reader, let us recall the data processing procedure from our previous work: the graph was constructed by first excluding non-English papers, entries missing key metadata (e.g., WoS unique identifier, title, year), or those with inconsistent self-citations. After cleaning, the resulting graph contained 964,811 nodes and 5,087,058 edges. The largest weakly connected component was retained, and degree-one nodes were removed, resulting in a graph of 722,623 nodes and 4,856,290 edges. For future extensions involving textual analysis, a filtered version including only papers with abstracts was created. This graph contains 698,135 nodes and 4,590,190 edges, with consecutively relabeled node IDs, stored in two files:
\begin{itemize}
    \item \texttt{abstr\_WOS\_2subj\_2000-2024.gph} (edge list)
    \item \texttt{abstr\_WOS\_2subj\_2000-2024\_nodes2uid.txt} (node-label list).
\end{itemize}
The overall structure of the graph is as follows:

\smallskip
 \noindent{\bf Paper and link distribution.} In the final graph, there are $505,\!613$ papers labeled ``M'', $192,\!340$ labeled ``OR\&MS'', and $182$ labeled with ``both'', representing papers in both categories. There are  $4,\!457,\!574$ internal links. Of these, $2,\!930,\!494$ are within the ``M'' category and $1,\!527,\!080$ within ``OR\&MS''. Notably, there are no internal links for the papers labeled ``both''. External links total $132,\!616$, with the majority ($131,\!186$) connecting ``M''~and~``OR\&MS''.
%

 \smallskip 
\noindent{\bf Degree distribution.} The degree distribution of the graph ranges from a minimum degree of $2$ to a maximum of $1436$. It is right-skewed, with the majority of nodes having low degrees -- over $60,\!000$ nodes have degree $2$, and the number decreases steadily with higher degrees. The mean degree is $13.15$, and the median is $9$, indicating a long tail of high-degree nodes. Table~\ref{tab:degree_distribution} compares the five lowest degrees (2–6), which account for $263,\!717$ nodes (37.79\% of total nodes), with the five highest-degree nodes. Although rare, these high-degree nodes play a key role in network reach, with their depth-2 neighborhoods containing respectively 6.67\% ($46,\!534$ nodes) and  8.05\% ($369,\!493$ edges) of total nodes and edges. Nearly the entire graph can be reached from these hubs at depth 8.
\begin{table}[!ht]
\centering
\caption{\small Degree distribution: Number of papers in 5 lowest degrees (left) and 5 highest degree papers and their neighborhood size of depth 2 (right). Percentages relative to total papers (698,135) and citation links (4,590,190).}\label{tab:degree_distribution}
\small
\begin{minipage}{0.35\linewidth}
\centering
\textbf{5 lowest degrees}
\begin{tabularx}{\linewidth}{>{\raggedright\arraybackslash}p{0.18\linewidth} >{\raggedleft\arraybackslash}X}
\toprule
\textbf{Degree} & \textbf{nodes} \\
\midrule
2 & 60,891 \\
3 & 57,227 \\
4 & 52,794 \\
5 & 48,473 \\
6 & 44,332 \\
\midrule
\textbf{Total} & \textbf{263,717}\\
\bottomrule
\end{tabularx}
\end{minipage}
\hfill
\begin{minipage}{0.55\linewidth}
\centering
\small
\textbf{5 highest degree papers}
\begin{tabularx}{\linewidth}{>{\raggedright\arraybackslash}p{1.8cm} >{\raggedleft\arraybackslash}X >{\raggedleft\arraybackslash}X >{\raggedleft\arraybackslash}X}
\toprule
\textbf{Paper ID} & \textbf{degree} & \textbf{nodes} & \textbf{edges} \\
\midrule
75    & 1,436 & 9,866  & 77,210 \\
6351  & 1,433 & 10,851 & 67,771 \\
2828  & 1,192 & 15,666 & 138,169 \\
5181  & 1,054 & 4,802  & 45,643 \\
278   &   937 & 5,349  & 40,700 \\
\midrule
\textbf{Total} & — & \textbf{46,534} & \textbf{369,493}\\
\bottomrule
\end{tabularx}
\end{minipage}
\end{table}

The structure and heterogeneity of this real-world citation graph present both challenges and opportunities for clustering. Its mix of dense cores and sparse peripheries makes it a suitable benchmark for evaluating community detection methods. In the following section, we apply spectral, Louvain, and Leiden clustering algorithms to investigate how well they capture the underlying organization of the network.

\section{Experiments and lessons learned}
The spectral, Louvain, and Leiden clustering algorithms were implemented in Python and executed on the same high-performance computing system equipped with two Intel Xeon Gold 6132 CPUs (28 cores, 56 threads, 2.60 GHz base, 3.7 GHz max) and 376~GiB of RAM. As a preliminary test, we applied the algorithms to a simple 7-node graph with a clear structure. All three methods produced consistent and reasonable results. This sanity check confirmed that our setup and code were functioning correctly.
 
We then applied the algorithms to the WoS citation graph described in the previous section. The spectral algorithm failed to produce a result within 24 hours, highlighting its limited scalability. In contrast, Louvain and Leiden returned results very fast (about 1 minute). These methods were implemented using the \texttt{igraph} and \texttt{leidenalg} libraries via the \texttt{community\_multilevel()} and \texttt{find\_partition()} functions, respectively. With default settings, both algorithms produced weakly structured clusterings. However, by tuning the resolution parameter in the Leiden algorithm, we obtained a more meaningful result at a resolution value of 0.05. In the following, we evaluate this outcome in detail.

\begin{figure}[!ht]
  \centering
  \includegraphics[width=\linewidth]{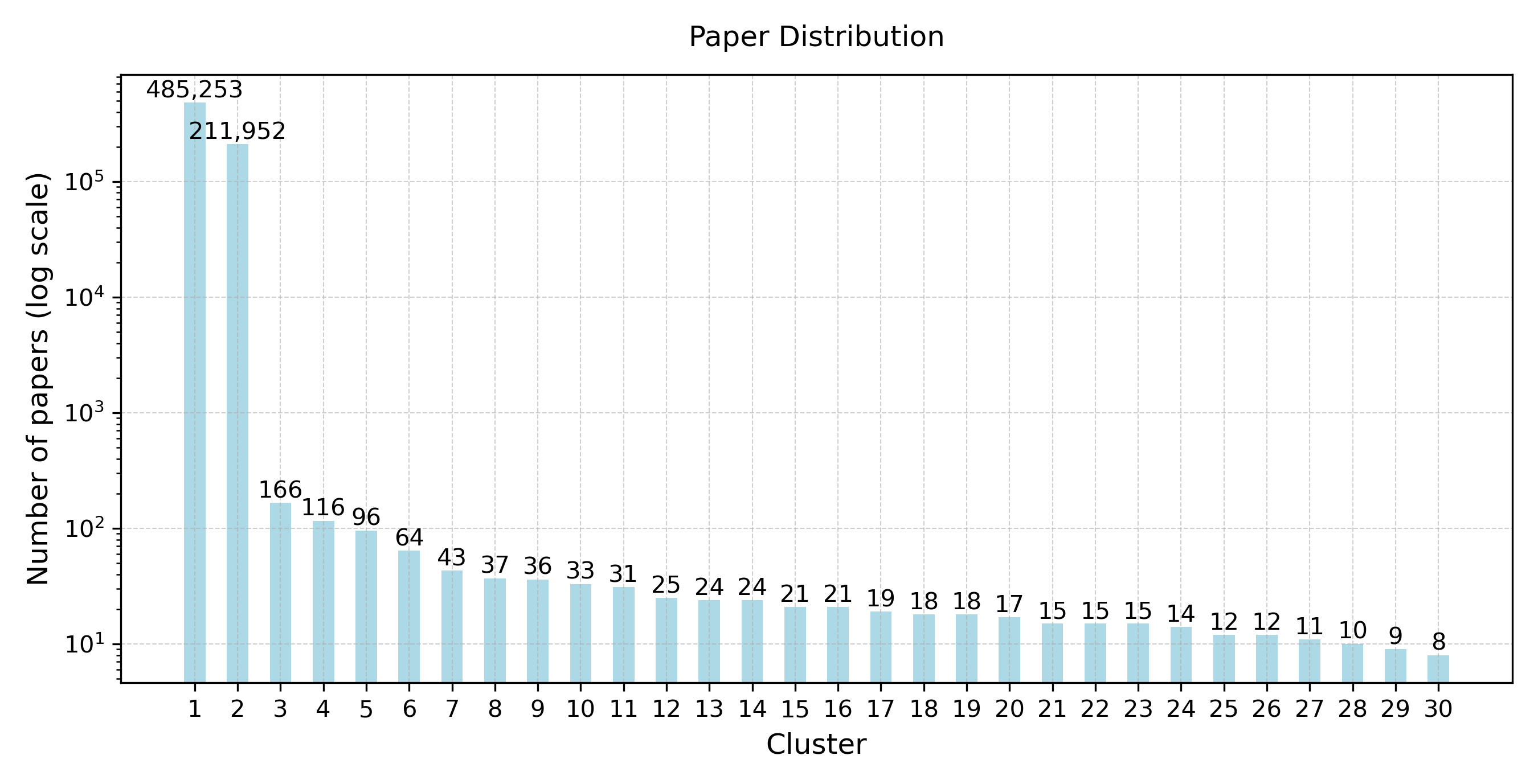}\caption{\small Solution by Leiden alg.: Paper distribution among clusters by size}
  \label{paper_distribution_pic}
\end{figure}

 \smallskip
\noindent{\bf Paper and link distribution.}
With the edge list as input, the Leiden algorithm with resolution 0.05 produced 30 clusters. As shown in Fig.~\ref{paper_distribution_pic}, the two largest clusters dominate: Cluster~1 has 485{,}253 nodes and Cluster~2 has 211{,}952, together comprising 697{,}205 papers (99.87\% of the total). The remaining 28 clusters are much smaller, ranging from 8 to 166 nodes.

\begin{figure}[!ht]
  \centering
  \includegraphics[width=\linewidth]{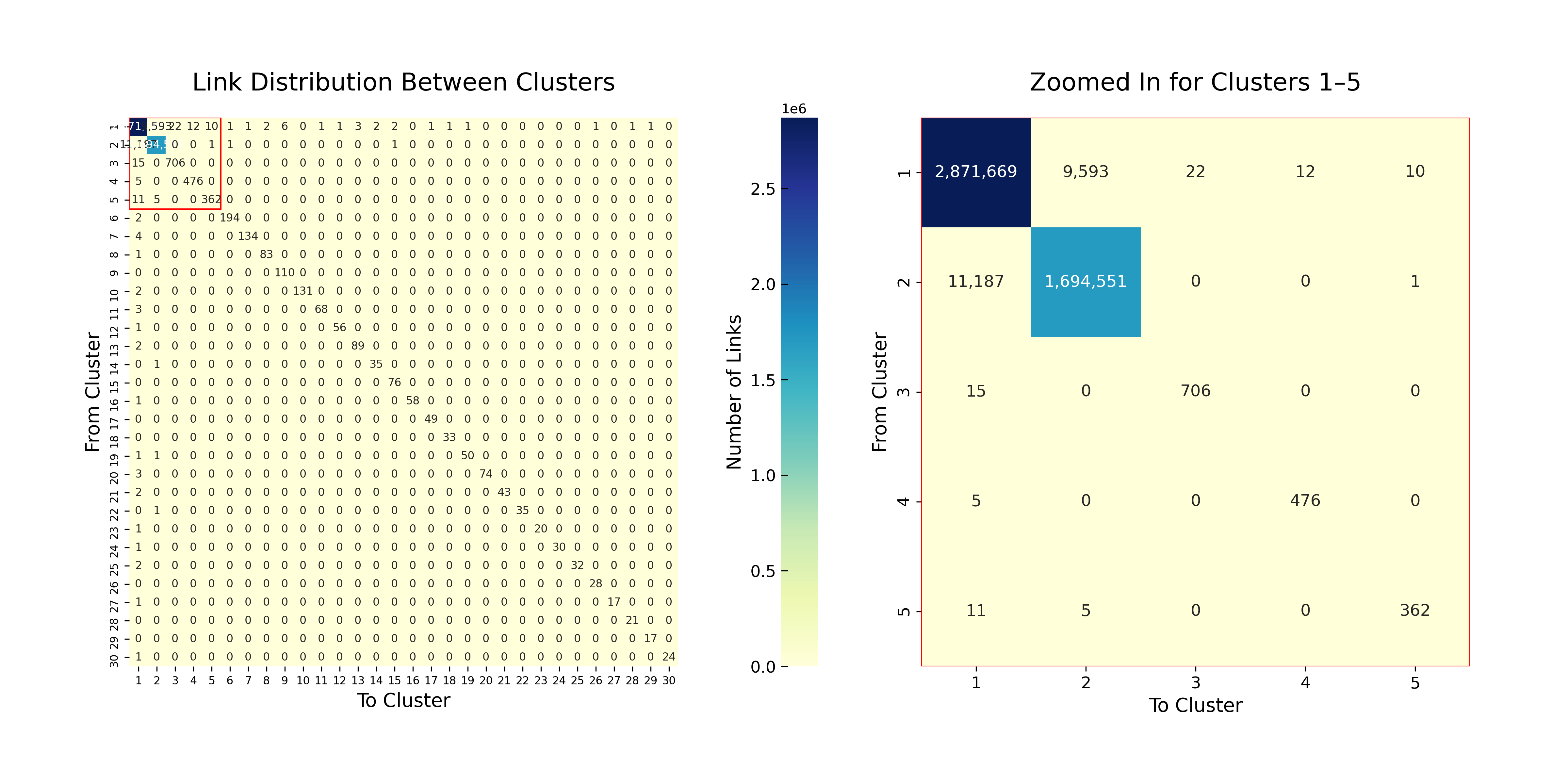}
  \caption{\small Solution by Leiden alg.: Heatmap of link distribution between clusters (left) and zoom-in for 5 biggest clusters (right)}
  \label{link_distribution_heatmap_pic}
\end{figure}

Fig.~\ref{link_distribution_heatmap_pic} shows a heatmap of link distribution across clusters (left) and a zoomed-in view of the top five (right). Of the total 4{,}590{,}190 links, 99.54\% (4{,}569{,}271) are within clusters and only 0.46\% (20{,}919) connect different clusters. Clusters~1 and~2 alone account for 99.93\% of all links -- 4{,}566{,}220 internal and 20{,}780 external -- indicating highly cohesive internal structures and minimal cross-cluster connectivity. Also notably, the remaining 28 smaller clusters have very few external connections and primarily link to Cluster 1 -- except Cluster 22, which links to Cluster 2. This suggests a natural simplification: merging Clusters~3–30 into Cluster~1 and Cluster~22 into Cluster~2 would yield a more coherent, label-aligned two-cluster structure.

To better understand the clustering solution versus the two known scientific fields, we will next use the label file to evaluate the fragmentation of the label and the purity of the cluster.

\noindent{\bf Label fragmentation.} To evaluate the coherence of clustering w.r.t the two known scientific fields, we computed the \textit{label fragmentation} metric. For each ground-truth label, we counted how many clusters its papers were assigned to and measured the concentration of these papers in the most dominant cluster.
\begin{table}[!ht]
\centering
\caption{\small Solution by Leiden alg.: Label fragmentation}\label{tab:label_fragmentation}
\resizebox{\textwidth}{!}{\input{./input/leiden_evaluation_label_fragmentation.tex}}
\end{table}

As shown in Table~\ref{tab:label_fragmentation}, the majority of nodes with the ``OR\&MS'' label (95.48\%) are concentrated in a single cluster (Cluster 2), indicating strong agreement with the clustering result. Similarly, the ``M'' label shows 94.24\% concentration in Cluster 1 but is distributed across 29 clusters in total, suggesting that while the clustering is effective at grouping similar nodes, some spread remains -- possibly due to topic granularity.

In contrast, the ``both'' label is more fragmented: only 58.79\% of its nodes fall in the dominant cluster. This reflects its interdisciplinary nature and the inherent challenge of assigning such nodes to a single community.

\begin{table}[!ht]
\centering
\caption{\small Solution by Leiden alg.: Cluster purity}\label{tab:cluster_purity}
\resizebox{0.7\textwidth}{!}{\input{./input/leiden_evaluation_cluster_purity.tex}}
\end{table}
\noindent{\bf Cluster purity.} To assess the semantic coherence of the detected clusters, we computed the \textit{purity} of each cluster w.r.t. ground-truth labels. For each cluster, we counted how many papers belonged to each label, identified the dominant label, and calculated purity as the fraction of papers with the dominant label.

Table~\ref{tab:cluster_purity} shows that Cluster~1 is highly pure (98.2\%) and strongly aligned with the ``M'' label. Similarly, Cluster~2 aligns with ``OR\&MS'' and achieves 86.6\% purity. All remaining clusters are small, mostly pure (often 100\%), and predominantly contain ``M'' nodes. Note that while ``M''-label nodes appear across 29 clusters, 28 of these can be merged with Cluster~1 due to their shared label and high purity, as observed about their lack of links to Cluster~2 in the link distribution matrix. 

The analysis on paper and link distribution, label fragmentation, and cluster purity demonstrates that the Leiden clustering solution has a clear structure and aligns pretty well with WoS domain labels.

\smallskip
\noindent{\bf Lessons learned.} Our exploratory experiments on this large, real-world citation graph from WoS highlight both the promise and limitations of commonly used clustering algorithms. Spectral clustering did not scale to the full dataset, making it less practical for networks of this size. In contrast, the Louvain and Leiden algorithms completed efficiently, making them more suitable choices when working with large graphs. While default settings yielded limited structure, basic tuning, such as adjusting the resolution parameter, helped reveal more interpretable patterns in the Leiden output. Although our analysis is still preliminary, these results highlight the value of fast, tunable algorithms as a starting point for deeper network exploration.

\section{Future work}
The initial findings suggest that further analysis is needed to understand the misalignment between the tuned Leiden clustering (resolution 0.05) and WoS labels, particularly for interdisciplinary or fragmented categories. The small isolated clusters (Clusters 3–30) may represent real subfields or result from data artifacts -- an open question worth exploring. Future work could also build on a strong hard partition as a foundation for soft clustering methods, which are better suited to the multi-topic nature of scientific publications. Finally, a high-quality clustering solution may support the reevaluation of field taxonomies.

 \medskip
\noindent{\bf Acknowledgements.} This work is co-funded by the European Union (European Regional Development Fund EFRE, Fund No.\! STIIV-001) and supported by the German Competence Network for Bibliometrics (Grant~No.\! 16WIK2101A).

\end{document}

%% file: input/leiden_evaluation_label_fragmentation.tex
\begin{tabular}{lrrrr}
\toprule
 \textbf{Label} &  \textbf{num. of clusters} &  \textbf{top cluster} &  \textbf{top cluster pct.} & \textbf{total papers} \\
\midrule
 OR\&MS &             5 &            2 &            95.48 &     192,340 \\
     M &            29 &            1 &            94.24 &     505,613 \\
  both &             3 &            1 &            58.79 &         182 \\
\bottomrule
\end{tabular}

%% file: input/leiden_evaluation_cluster_purity.tex
\begin{tabular}{lrrrcr}
\toprule
\textbf{Cluster} &       \textbf{M} &   \textbf{OR\&MS} &  \textbf{both} & \textbf{dominant label} &    \textbf{purity} \\
\midrule
1  &  476.466 &    8.680 &   107 &              M &  0.981892 \\
2  &   28.238 &  183.640 &    74 &          OR\&MS &  0.866423 \\
3  &     166 &       0 &     0 &              M &  1.000000 \\
4  &     116 &       0 &     0 &              M &  1.000000 \\
5  &      92 &       4 &     0 &              M &  0.958333 \\
6  &      63 &       0 &     1 &              M &  0.984375 \\
7  &      43 &       0 &     0 &              M &  1.000000 \\
8  &      37 &       0 &     0 &              M &  1.000000 \\
9  &      36 &       0 &     0 &              M &  1.000000 \\
10 &      33 &       0 &     0 &              M &  1.000000 \\
11 &      31 &       0 &     0 &              M &  1.000000 \\
12 &      25 &       0 &     0 &              M &  1.000000 \\
13 &      24 &       0 &     0 &              M &  1.000000 \\
14 &      23 &       1 &     0 &              M &  0.958333 \\
15 &      21 &       0 &     0 &              M &  1.000000 \\
16 &      21 &       0 &     0 &              M &  1.000000 \\
17 &      19 &       0 &     0 &              M &  1.000000 \\
18 &      18 &       0 &     0 &              M &  1.000000 \\
19 &      18 &       0 &     0 &              M &  1.000000 \\
20 &      17 &       0 &     0 &              M &  1.000000 \\
21 &      15 &       0 &     0 &              M &  1.000000 \\
23 &      15 &       0 &     0 &              M &  1.000000 \\
24 &      14 &       0 &     0 &              M &  1.000000 \\
25 &      12 &       0 &     0 &              M &  1.000000 \\
26 &      12 &       0 &     0 &              M &  1.000000 \\
27 &      11 &       0 &     0 &              M &  1.000000 \\
28 &      10 &       0 &     0 &              M &  1.000000 \\
29 &       9 &       0 &     0 &              M &  1.000000 \\
30 &       8 &       0 &     0 &              M &  1.000000 \\
22 &       0 &      15 &     0 &          OR\&MS &  1.000000 \\
\bottomrule
\end{tabular}

%% file: gor_main.bbl
\begin{thebibliography}{99}

\bibitem{Blondel_2008}
V.D. Blondel, J.L. Guillaume, R. Lambiotte, E. Lefebvre:  
Fast unfolding of communities in large networks,  
Journal of Statistical Mechanics: Theory and Experiment, 10, P10008, 2008.

\bibitem{Traag_2019}
V.A. Traag, L. Waltman, N.J. van Eck: From Louvain to Leiden: guaranteeing well-connected communities,  
Scientific Reports, 9, 5233, 2019.

\bibitem{Ng_2002}
A.Y. Ng, M.I. Jordan, Y. Weiss: On spectral clustering: analysis and an algorithm,  
NeurIPS, 2002.






\bibitem{webofscience}
Clarivate Analytics, Web of Science, \href{https://www.webofscience.com}{https://www.webofscience.com}

\bibitem{Huong_Ida_Koch} V.T. Huong, I. Litzel, T. Koch: Similarity-based fuzzy clustering scientific articles: potentials and challenges from mathematical and computational perspectives, 2025. 


\end{thebibliography}
